\documentclass[10pt,a4paper,onecolumn]{article}
\usepackage{marginnote}
\usepackage{graphicx}
\usepackage{xcolor}
\usepackage{authblk,etoolbox}
\usepackage{titlesec}
\usepackage{calc}
\usepackage{tikz}
\usepackage{hyperref}
\hypersetup{colorlinks,breaklinks,
            urlcolor=[rgb]{0.0, 0.5, 1.0},
            linkcolor=[rgb]{0.0, 0.5, 1.0}}
\usepackage{caption}
\usepackage{tcolorbox}
\usepackage{amssymb,amsmath}
\usepackage{ifxetex,ifluatex}
\usepackage{seqsplit}
\usepackage{xstring}

\let\textttOrig=\texttt
\def\texttt#1{\expandafter\textttOrig{\seqsplit{#1}}}
\renewcommand{\seqinsert}{\ifmmode
  \allowbreak
  \else\penalty6000\hspace{0pt plus 0.02em}\fi}

\makeatletter
\let\href@Orig=\href
\def\href@Urllike#1#2{\href@Orig{#1}{\begingroup
    \def\Url@String{#2}\Url@FormatString
    \endgroup}}
\def\href@Notdoi#1#2{\def\tempa{#1}\def\tempb{#2}%
  \ifx\tempa\tempb\relax\href@Urllike{#1}{#2}\else
  \href@Orig{#1}{#2}\fi}
\def\href#1#2{%
  \IfBeginWith{#1}{https://doi.org}%
  {\href@Urllike{#1}{#2}}{\href@Notdoi{#1}{#2}}}
\makeatother

\usepackage[top=3.5cm, bottom=3cm, right=1.5cm, left=1.5cm, headheight=2.2cm, reversemp, includemp, marginparwidth=4.3cm]{geometry}

\titleformat{\section}
  {\normalfont\sffamily\Large\bfseries}
  {}{0pt}{}
\titleformat{\subsection}
  {\normalfont\sffamily\large\bfseries}
  {}{0pt}{}
\titleformat{\subsubsection}
  {\normalfont\sffamily\bfseries}
  {}{0pt}{}
\titleformat*{\paragraph}
  {\sffamily\normalsize}

\usepackage{fancyhdr}
\makeatletter
\let\ps@plain\ps@fancy
\fancyheadoffset[L]{4.5cm}
\fancyfootoffset[L]{4.5cm}

\definecolor{linky}{rgb}{0.0, 0.5, 1.0}

\newtcolorbox{repobox}
   {colback=red, colframe=red!75!black,
     boxrule=0.5pt, arc=2pt, left=6pt, right=6pt, top=3pt, bottom=3pt}

\newcommand{\ExternalLink}{%
   \tikz[x=1.2ex, y=1.2ex, baseline=-0.05ex]{%
       \begin{scope}[x=1ex, y=1ex]
           \clip (-0.1,-0.1)
               --++ (-0, 1.2)
               --++ (0.6, 0)
               --++ (0, -0.6)
               --++ (0.6, 0)
               --++ (0, -1);
           \path[draw,
               line width = 0.5,
               rounded corners=0.5]
               (0,0) rectangle (1,1);
       \end{scope}
       \path[draw, line width = 0.5] (0.5, 0.5)
           -- (1, 1);
       \path[draw, line width = 0.5] (0.6, 1)
           -- (1, 1) -- (1, 0.6);
       }
   }

\patchcmd{\@maketitle}{center}{flushleft}{}{}
\patchcmd{\@maketitle}{center}{flushleft}{}{}
\patchcmd{\@maketitle}{\LARGE}{\LARGE\sffamily}{}{}
\def\maketitle{{%
  
  \AB@maketitle}}
\makeatletter
\renewcommand\AB@affilsepx{ \protect\Affilfont}
\renewcommand\AB@affilnote[1]{{\bfseries #1}\hspace{3pt}}
\renewcommand{\affil}[2][]%
   {\newaffiltrue\let\AB@blk@and\AB@pand
      \if\relax#1\relax\def\AB@note{\AB@thenote}\else\def\AB@note{#1}%
        \setcounter{Maxaffil}{0}\fi
        \begingroup
        \let\href=\href@Orig
        \let\texttt=\textttOrig
        \let\protect\@unexpandable@protect
        \def\thanks{\protect\thanks}\def\footnote{\protect\footnote}%
        \@temptokena=\expandafter{\AB@authors}%
        {\def\\{\protect\\\protect\Affilfont}\xdef\AB@temp{#2}}%
         \xdef\AB@authors{\the\@temptokena\AB@las\AB@au@str
         \protect\\[\affilsep]\protect\Affilfont\AB@temp}%
         \gdef\AB@las{}\gdef\AB@au@str{}%
        {\def\\{, \ignorespaces}\xdef\AB@temp{#2}}%
        \@temptokena=\expandafter{\AB@affillist}%
        \xdef\AB@affillist{\the\@temptokena \AB@affilsep
          \AB@affilnote{\AB@note}\protect\Affilfont\AB@temp}%
      \endgroup
       \let\AB@affilsep\AB@affilsepx
}
\makeatother

\renewcommand\Affilfont{\sffamily\small\mdseries}
\ifnum 0\ifxetex 1\fi\ifluatex 1\fi=0 
  \usepackage[T1]{fontenc}
  \usepackage[utf8]{inputenc}

\else 
  \ifxetex
    \usepackage{mathspec}
  \else
    \usepackage{fontspec}
  \fi
  \defaultfontfeatures{Ligatures=TeX,Scale=MatchLowercase}

\fi
\IfFileExists{upquote.sty}{\usepackage{upquote}}{}
\IfFileExists{microtype.sty}{%
\usepackage{microtype}
\UseMicrotypeSet[protrusion]{basicmath} 
}{}

\usepackage{hyperref}
\hypersetup{unicode=true,
            pdftitle={10.21105.joss.02554.pdf},
            pdfborder={0 0 0},
            breaklinks=true}
\let\addcontentslineOrig=\addcontentsline
\def\addcontentsline#1#2#3{\bgroup
  \let\texttt=\textttOrig\addcontentslineOrig{#1}{#2}{#3}\egroup}
\let\markbothOrig\markboth
\def\markboth#1#2{\bgroup
  \let\texttt=\textttOrig\markbothOrig{#1}{#2}\egroup}
\let\markrightOrig\markright
\def\markright#1{\bgroup
  \let\texttt=\textttOrig\markrightOrig{#1}\egroup}

\usepackage{graphicx,grffile}
\makeatletter
\def\maxwidth{\ifdim\Gin@nat@width>\linewidth\linewidth\else\Gin@nat@width\fi}
\def\maxheight{\ifdim\Gin@nat@height>\textheight\textheight\else\Gin@nat@height\fi}
\makeatother
\setkeys{Gin}{width=\maxwidth,height=\maxheight,keepaspectratio}
\IfFileExists{parskip.sty}{%
\usepackage{parskip}
}{
\setlength{\parindent}{0pt}
\setlength{\parskip}{6pt plus 2pt minus 1pt}
}
\ifx\paragraph\undefined\else
\let\oldparagraph\paragraph
\renewcommand{\paragraph}[1]{\oldparagraph{#1}\mbox{}}
\fi
\ifx\subparagraph\undefined\else
\let\oldsubparagraph\subparagraph
\renewcommand{\subparagraph}[1]{\oldsubparagraph{#1}\mbox{}}
\fi

\title{The Dusty Evolved Star Kit (DESK): A Python package for fitting the Spectral Energy Distribution of Evolved Stars}

        \author[1]{Steven R. Goldman}

      \affil[1]{Space Telescope Science Institute, 3700 San Martin Drive, Baltimore, MD 21218, USA}
  \date{\vspace{-5ex}}

\begin{document}
\maketitle

\marginpar{
  \sffamily\small

  {\bfseries DOI:} \href{https://joss.theoj.org/papers/10.21105/joss.02554}{\color{linky}{10.21105/joss.02554}}

  \vspace{2mm}

  {\bfseries Software}
  \begin{itemize}
    \setlength\itemsep{0em}
    \item \href{https://github.com/openjournals/joss-reviews/issues/2554}{\color{linky}{Review}} \ExternalLink
    \item \href{https://github.com/s-goldman/Dusty-Evolved-Star-Kit}{\color{linky}{Repository}} \ExternalLink
    \item \href{https://doi.org/10.5281/zenodo.4064241}{\color{linky}{Archive}} \ExternalLink
  \end{itemize}

  \noindent\rule{4cm}{0.4pt}
  \medskip

  {\bfseries Editor:} \href{https://www.arfon.org/}{\color{linky}{Arfon Smith}} \ExternalLink \\ 
  {\bfseries Reviewers:}
  \begin{itemize}
    \setlength\itemsep{0em}
    \item \href{https://github.com/Deech08}{\color{linky}{@Deech08}}
    \item \href{https://github.com/TomGoffrey}{\color{linky}{@TomGoffrey}
}
  \end{itemize}

  \medskip

  {\bfseries Submitted:} 24 July 2020\\
  {\bfseries Published:} 02 October 2020

  \vspace{2mm}
  {\bfseries License}\\
  Authors of papers retain \\
  copyright and release the work \\
  under a Creative Commons \\
  Attribution 4.0 International \\
  License (\href{https://creativecommons.org/licenses/by/4.0/}{\color{linky}{CC BY 4.0}}).
}

\newcommand{\cmfont}{\usefont\encodingdefault
  \sfdefault
  \seriesdefault
  \shapedefault
  \relax}
\cmfont
\vspace{0.5cm}

\hypertarget{summary}{%
\section{Summary}\label{summary}}

One of the few ways that we can understand the environment around dusty stars and how much material they contribute back to the Universe, is by fitting their brightness at different wavelengths with models that account for how the energy transfers through the dust. Codes for creating models have been developed and refined (Elitzur \& Ivezić, \href{https://doi.org/10.1046/j.1365-8711.2001.04706.x}{2001}; Ueta \& Meixner, \href{https://doi.org/10.1086/367818}{2003}), but a code for easily fitting data to grids of realistic models has been up-to-this-point unavailable.

The DESK is a python package designed to compare the best fits of different stellar samples and model grids for a better understanding of the results and their uncertainties. The package fits the Spectral Energy Distribution (SED) of evolved stars, using photometry or spectra, to grids of radiative transfer models using a least-squares method. The package includes newly created grids using a variety of different dust species, and state-of-the-art dust growth grids (Nanni et al., \href{https://doi.org/10.1093/mnras/stx2601}{2019}). Early versions of the code have been used in (Goldman et al., \href{https://doi.org/10.3847/1538-4357/ab418a}{2019}; Goldman et al., \href{https://doi.org/10.1093/mnras/stx2601}{2018}, \href{https://doi.org/10.1093/mnras/stw2708}{2017}; Orosz et al., \href{https://doi.org/10.3847/1538-3881/153/3/119}{2017})

\section{Statement of need}
To understand the ranges and estimated errors of fitted results, they must be compared to results from different model grids. Results from these grids (e.g. luminosity, mass-loss rate) can vary dramatically as a result of the unknown dust properties and geometry of evolved stars (Sargent et al., \href{https://doi.org/10.1088/0004-637X/716/1/878}{2010}; Srinivasan, Sargent, \& Meixner, \href{https://doi.org/10.1051/0004-6361/201117033}{2011}; Wiegert, Groenewegen, Jorissen, Decin, \& Danilovich, \href{{http://arxiv.org/abs/2008.11525}}{2020}). This is especially true of the oxygen-rich Asymptotic Giant Branch (AGB) stars. Adding to this challenge is the fact that models are calculated based on measured values of the dust (optical constants) which can not be interpolated over. A robust method for testing different model grids will be particularly important given the wealth of infrared data to come from the James Webb Space Telescope (JWST).

\section{User interface}
The package can be installed using {\rm pip} and imported within python. Using “entrypoints”, the package can also be accessed from any terminal prompt once installed. The fitting method uses a brute-force technique to ensure a true best fit. New grids of multi-dimensional radiative transfer models will be added to the model grid library as they are developed. The available model grids for this version are listed in Table 1.

\begin{figure}
\section*{Figures}\label{figures}
\centering
\includegraphics[width=0.7\columnwidth]{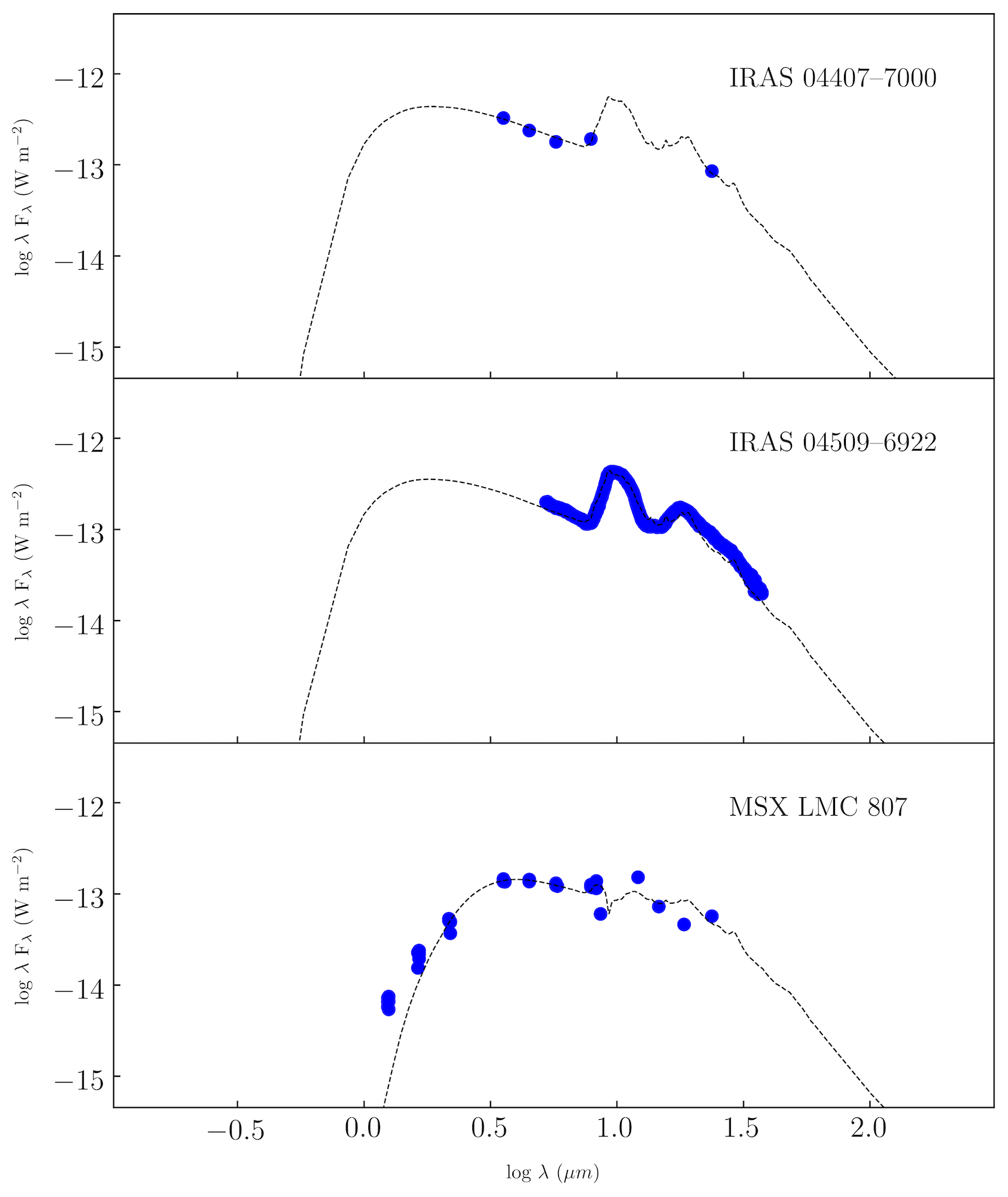}
\caption{An example of three massive oxygen-rich AGB stars in the Large Magellanic Cloud (LMC) galaxy fit with the default oxygen-rich model grid (Oss-Orich-bb). These three example sources can be fit, and this figure can be created, using the command {\rm \textbf{desk fit}} and then the command {\rm \textbf{desk sed}}.}
\end{figure}
\begin{figure}
\centering
\includegraphics[width=\columnwidth]{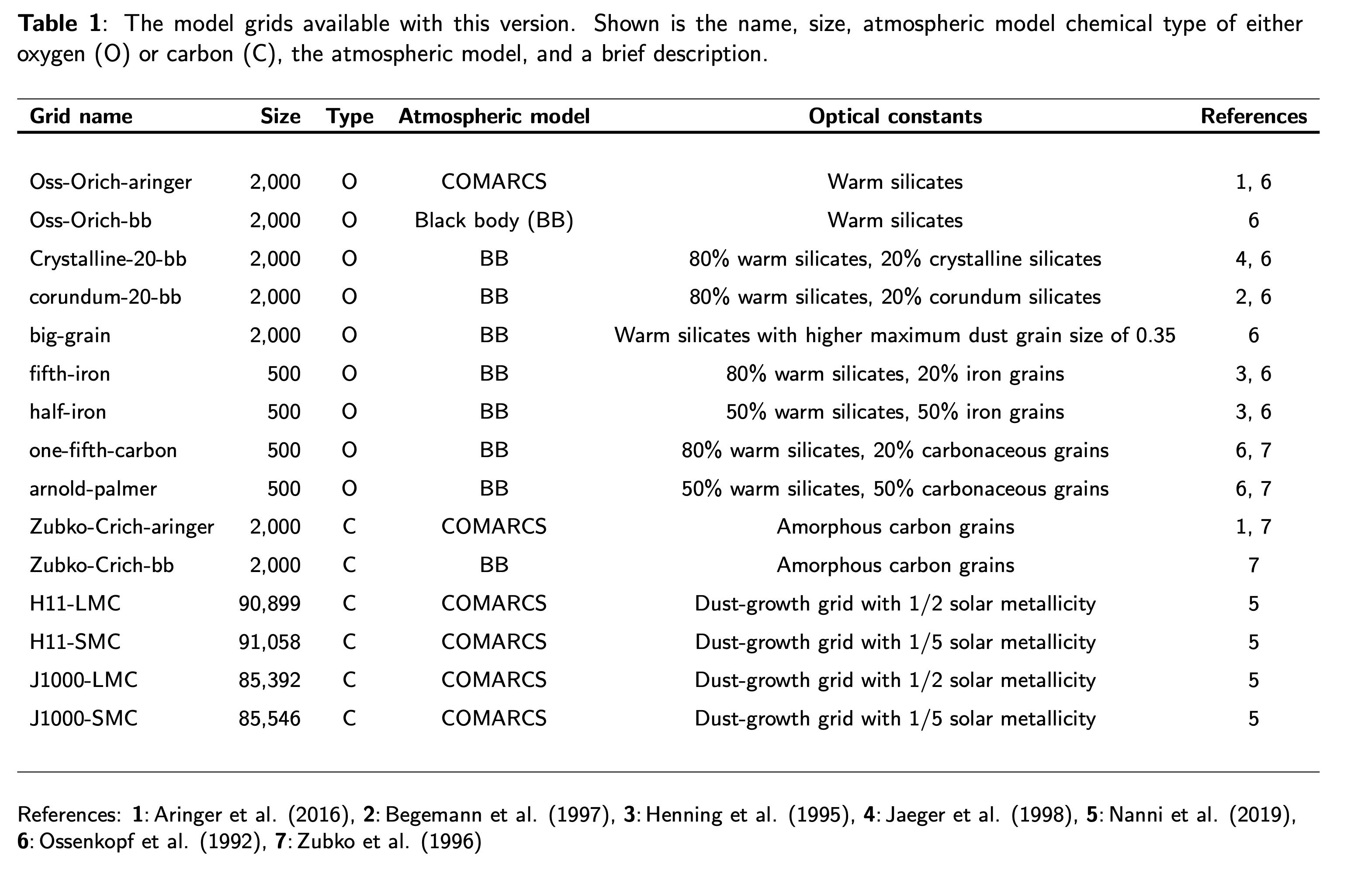}
\end{figure}

\hypertarget{references}{%
\section*{References}\label{references}}
\addcontentsline{toc}{section}{References}

\begin{list}{}{\leftmargin=1.5em \itemindent=-1.5em \itemsep=-1pt}

\item Aringer, B., Girardi, L., Nowotny, W., Marigo, P., \& Bressan, A. (2016). Synthetic photometry for M and K giants and stellar evolution: hydrostatic dust-free model atmospheres and chemical abundances, 457(4), 3611–3628. doi:\href{https://doi.org/10.1093/mnras/stw222}{\color{linky}{10.1093/mnras/stw222}}

\item Begemann, B., Dorschner, J., Henning, T., Mutschke, H., Gürtler, J., Kömpe, C., \& Nass, R. (1997). Aluminum Oxide and the Opacity of Oxygen-rich Circumstellar Dust in the 12-17 Micron Range, 476(1), 199–208. doi:\href{https://iopscience.iop.org/article/10.1086/303597}{\color{linky}{10.1086/303597}}

\item Elitzur, M., \& Ivezi\'{c}, \v{Z}. (2001). Dusty winds - I. Self-similar solutions, 327, 403–421. doi:\href{https://doi.org/10.1046/j.1365-8711.2001.04706.x}{10.1046/j.1365-8711.2001.04706.x}

\item Goldman, S. R., Boyer, M. L., McQuinn, K. B., Sloan, G. C., McDonald, I., Loon, J. T. van, Zijlstra, A. A., et al. (2019). AGB stars in the nearby dwarf galaxy: Leo p. ApJ, 884, 152. doi:\href{https://doi.org/10.3847/1538-4357/ab418a}{10.3847/1538-4357/ab418a}

\item Goldman, S. R., van Loon, J. T., Gómez, J. F., Green, J. A., Zijlstra, A. A., Nanni, A., Imai, H., et al. (2018). A dearth of OH/IR stars in the Small Magellanic Cloud, 473, 3835–3853. doi:\href{https://doi.org/10.1093/mnras/stx2601}{10.1093/mnras/stx2601}

\item Goldman, S. R., van Loon, J. T., Zijlstra, A. A., Green, J. A., Wood, P. R., Nanni, A., Imai, H., et al. (2017). The wind speeds, dust content, and mass-loss rates of evolved AGB and RSG stars at varying metallicity, 465(1), 403–433. doi:\href{https://doi.org/10.1093/mnras/stw2708}{10.1093/mnras/stw2708}

\item Henning, T., Begemann, B., Mutschke, H., \& Dorschner, J. (1995). Optical properties of oxide dust grains., 112, 143.

\item Jaeger, C., Molster, F. J., Dorschner, J., Henning, T., Mutschke, H., \& Waters, L. B. F. M. (1998). Steps toward interstellar silicate mineralogy. IV. The crystalline revolution, 339, 904–916.

\item Nanni, A., Groenewegen, M. A. T., Aringer, B., Rubele, S., Bressan, A., Loon, J. T. van, Goldman, S. R., et al. (2019). The mass-loss, expansion velocities, and dust production rates of carbon stars in the magellanic clouds, 487, 502–521. doi:\href{https://doi.org/10.1093/mnras/stz1255}{10.1093/mnras/stz1255}

\item Orosz, G., Imai, H., Dodson, R., Rioja, M. J., Frey, S., Burns, R. A., Etoka, S., et al. (2017). Astrometry of OH/IR Stars Using 1612 MHz Hydroxyl Masers. I. Annual Parallaxes of WX Psc and OH138.0+7.2, 153. doi:\href{https://doi.org/10.3847/1538-3881/153/3/119}{10.3847/1538-3881/153/3/119}

\item Ossenkopf, V., Henning, T., \& Mathis, J. S. (1992). Constraints on cosmic silicates., 261, 567–578.

\item Sargent, B. A., Srinivasan, S., Meixner, M., Kemper, F., Tielens, A. G. G. M., Speck, A. K., Matsuura, M., et al. (2010). The mass-loss return from evolved stars to the large magellanic cloud. II. Dust properties for oxygen-rich asymptotic giant branch stars, 716, 878–890. doi:\href{https://doi.org/10.1088/0004-637X/716/1/878}{10.1088/0004-637X/716/1/878}

\item Srinivasan, S., Sargent, B. A., \& Meixner, M. (2011). The mass-loss return from evolved stars to the Large Magellanic Cloud. V. The GRAMS carbon-star model grid, 532, A54. doi:\href{https://doi.org/10.1051/0004-6361/201117033}{10.1051/0004-6361/201117033}

\item Ueta, T., \& Meixner, M. (2003). 2-dust: A dust radiative transfer code for an axisymmetric system, 586, 1338–1355. doi:\href{https://doi.org/10.1086/367818}{10.1086/367818}

\item Wiegert, J., Groenewegen, M. A. T., Jorissen, A., Decin, L., \& Danilovich, T. (2020). How to disentangle geometry and mass-loss rate from AGB-star spectral energy distributions – The case of EP Aqr. arXiv e-prints, arXiv:2008.11525. Retrieved from \url{http://arxiv.org/abs/2008.11525}

\item Zubko, V. G., Mennella, V., Colangeli, L., \& Bussoletti, E. (1996). Optical constants of cosmic carbon analogue grains - I. Simulation of clustering by a modified continuous distribution of ellipsoids, 282(4), 1321–1329. doi:\href{https://doi.org/10.1093/mnras/282.4.1321}{10.1093/mnras/282.4.1321}
\end{list}
\end{document}